\documentclass[preprint]{article}

\usepackage[nonatbib]{neurips_2024}

\usepackage[utf8]{inputenc} 
\usepackage[T1]{fontenc}    
\usepackage[hyphens]{url}   
\usepackage{hyperref}       
\usepackage{url}            
\usepackage{booktabs}       
\usepackage{amsfonts}       
\usepackage{nicefrac}       
\usepackage{microtype}      
\usepackage{xcolor}         

\usepackage{float} 
\usepackage{graphicx}
\usepackage{amsmath}
\usepackage{subcaption}
\usepackage{setspace}

\usepackage{authblk}  

\usepackage{xcolor}

\usepackage[numbers]{natbib}  
\newcommand{\verify}[2][]{{\color{red} #2}}

\makeatletter
\renewcommand{\@noticestring}{}  
\makeatother

\title{Improving Diseases Predictions Utilizing External Bio-Banks}

\author[1]{Hido Pinto}
\author[1,2]{Eran Segal}

\affil[1]{Weizmann Institute of Science}
\affil[2]{Mohamed Bin Zayed University of Artificial Intelligence}

\affil[ ]{\texttt{yehudahido.pinto@weizmann.ac.il, eran.segal@weizmann.ac.il}}

\date{}  

\begin{document}

\maketitle

\begin{abstract}\label{abstract_bio_bank}


Machine learning has been successfully used in critical domains, such as medicine. 
However, extracting meaningful insights from biomedical data is often constrained by the lack of their available disease labels.
In this research, we demonstrate how machine learning can be leveraged to enhance explainability and uncover biologically meaningful associations, even when predictive improvements in disease modeling are limited. 
We train LightGBM models from scratch on our dataset (10K) to impute metabolomics features and apply them to the UK Biobank (UKBB) for downstream analysis.
The imputed metabolomics features are then used in survival analysis to assess their impact on disease-related risk factors.

As a result, our approach successfully identified biologically relevant connections that were not previously known to the predictive models. 
Additionally, we applied a genome-wide association study (GWAS) on key metabolomics features, revealing a link between vascular dementia and smoking. Although being a well-established epidemiological relationship, this link was not embedded in the model's training data, which validated the method's ability to extract meaningful signals. 
Furthermore, by integrating survival models as inputs in the 10K data, we uncovered associations between metabolic substances and obesity, demonstrating the ability to infer disease risk for future patients without requiring direct outcome labels.

These findings highlight the potential of leveraging external bio-banks to extract valuable biomedical insights, even in data-limited scenarios. Our results demonstrate that machine learning models trained on smaller datasets can still be used to uncover real biological associations when carefully integrated with survival analysis and genetic studies.

\end{abstract}

\section{Introduction} \label{bio_bank:introduction}

Metabolomics, the large-scale study of small molecules in biological systems, has its roots in analytical chemistry and systems biology. It is emerging as a powerful tool in healthcare, offering insights into disease mechanisms for both predictive modeling and clinical research \cite{lgbm_metabs, metabs_cardio}.
However, due to the high cost and complexity of metabolomics profiling, they are often not collected. Even large-scale datasets such as the UK Biobank (UKBB), with its significant amount of resources, takes a long time filling in metabolomics measurements for all participants while focusing on a single method(nmr) \cite{ukbb_metabs}. 
To address this limitation, machine learning models can be trained on smaller but richer datasets to impute missing metabolomics features, enabling downstream analyses.

In this study, we explore the potential of metabolomics imputation to improve disease prediction and enhance biological insights. By leveraging machine learning models trained on smaller datasets, we assess their ability to predict missing metabolomics features and integrate them into survival analysis and genetic studies. Our findings reveal biologically meaningful associations, highlighting the potential of external bio-banks in predictive modeling.

\section{Related Work} \label{bio_bank:related_work}

The integration of machine learning with biomedical datasets has been widely explored for both prediction and explainability in disease modeling. 
Gradient boosting methods such as LightGBM \cite{lgbm_paper} have been successfully applied to metabolic trait prediction \cite{lgbm_metabs}, while survival analysis models such as CoxPH \cite{Cox1972} have been successfully applied to disease risk assessment \cite{cox_ph_disease}. 
However, the use of machine learning to impute metabolomics features for downstream survival analysis remains an emerging area of research.

Several studies have investigated the impact of imputed omics data on disease prediction. 
For example, references \cite{classic_metabs_imp, deep_metabs_imp} proposed classical and deep learning-based imputation frameworks for missing metabolomics profiles in large-scale cohorts, demonstrating improvements in feature coverage but reporting mixed results in downstream predictive tasks. 
Similarly, the research in \cite{metabs_cardio} explored the use of proteomics and metabolomics imputations for cardiovascular disease risk modeling, finding that while direct improvements in prediction were modest, the imputed features provided valuable biological context.

In addition, combining machine learning-derived features with genome-wide association studies (GWAS) has proven useful for discovering new disease-related genetic links. 
Previous research has leveraged GWAS to validate metabolomics-based predictions \cite{gwas_metabs_val}, demonstrating how external datasets can be used to reinforce the interpretability of machine learning models.

Within the SegalLab, a prior study explored a related concept by training survival models on UKBB and transferring them into 10K data to generate synthetic disease labels for validation. 
While this work has yet to be published, it serves as an important precursor to our study, reinforcing the value of cross-dataset modeling. 
This research extends that approach by leveraging metabolomics imputation in UKBB to uncover well-established biological insights, such as vascular dementia and smoking association \cite{dementia_smoking, dementia_smoking_2} from GWAS, and identifying metabolic pathways associated with obesity \cite{hers_phd, solphat_nature} without requiring labeled disease outcomes.

These findings highlight the broader potential of using external bio-banks not just for disease prediction but also for extracting meaningful biomedical insights, even in data-limited scenarios.

\section{Data}

During the project we used 2 different Bio-Banks - 10K \cite{shilo2021cohort} and UKBB \cite{ukbb_paper}.

\medskip
\noindent
10K - Prof. Segal's main Bio-Bank, contains diverse medical data on 10,000 subjects over repeated visits(every 2 years) in Israel.

\medskip
\noindent
UKBB - A large an known Bio-Bank, contains diverse medical data on ~500,000 subjects from the UK with some of them returning for more visits.

\noindent
Being relatively young, at this moment 10K Bio-Bank lacks numbers to function as a many diseases cohort.
Thus we used UKBB for their multiple cohorts.

\bigskip
\noindent
Both Bio-Banks has multiple data modalities in common such as: Body measures, Blood test results, and life-style questionnaires.
These features will be referred to as the \textbf{shared features}.

\noindent
10K also contains a few metabolomics features, one of them will be simply referred to as \textbf{metabolomics}, the other, which will be referred to as \textbf{nmr} stand for metabolomics data collected through a NMR(Nuclear Magnetic Resonance) method. Another 10K data modality being used during this project is Dexa(Dual-energy X-ray Absorptiometry) which was used to get bone density and body composition features.

\noindent
Similar to 10K UKBB also contains nmr metabolomics gathered in the same method, and Dexa data gathered in different methods.

\section{Methods} \label{bio_bank:methods}

\subsection{LightGBM} \label{bio_bank:methods:lgbm}
Light Gradient Boosting Machine (LightGBM) is a gradient boosting framework that utilizes decision trees for classification and regression tasks. 
It is particularly well-suited for handling large datasets due to its histogram-based learning approach, which significantly reduces memory usage and computation time. 
Unlike traditional boosting algorithms that grow trees level-wise, LightGBM employs a leaf-wise growth strategy, leading to faster convergence and improved performance.  

LightGBM effectively processes missing values and categorical features through its optimized histogram-based learning approach.
Additionally, it supports parallel and GPU-based training, making it scalable for high-dimensional data. 
These properties make LightGBM a strong candidate for predictive modeling, including applications in survival analysis when combined with appropriate loss functions.

\subsection{CoxPH Model} \label{bio_bank:methods:cox_ph}
The Cox Proportional Hazards (CoxPH) model is a widely used statistical method for survival analysis. 
It estimates the effect of covariates on the hazard function, which represents the instantaneous risk of an event occurring at a given time. The model assumes that the hazard function for an individual $i$ is given by:

\begin{equation}
    h_i(t) = h_0(t) \exp(\beta_1 x_{i1} + \beta_2 x_{i2} + \dots + \beta_p x_{ip}),
\end{equation}

where $h_0(t)$ is the baseline hazard function, $\beta_j$ are the regression coefficients, and $x_{ij}$ are the covariates. 
A key assumption of the CoxPH model is the proportional hazards assumption, which states that the hazard ratios between individuals remain constant over time.

CoxPH is advantageous due to its semi-parametric nature, which does not require specification of the baseline hazard function. 
However, it can be limited in cases where the proportional hazards assumption is violated or when handling high-dimensional data.

\subsection{Survival Analysis} \label{bio_bank:methods:survival}
Survival analysis is a statistical framework for modeling time-to-event data, where the objective is to estimate the probability of an event (e.g., failure, death, or disease progression) occurring over time. 
A fundamental aspect of survival analysis is the presence of censored data, where the exact event time is unknown for some observations due to study limitations or loss to follow-up.

Key concepts in survival analysis include the survival function:

\begin{equation}
    S(t) = P(T > t),
\end{equation}

which represents the probability that the event has not yet occurred by time $t$, and the hazard function:

\begin{equation}
    h(t) = \lim_{\Delta t \to 0} \frac{P(t \leq T < t + \Delta t \mid T \geq t)}{\Delta t},
\end{equation}

which describes the instantaneous risk of the event occurring at time $t$.

Common approaches in survival analysis include the Kaplan-Meier estimator for non-parametric survival estimation and the CoxPH model for regression-based analysis. 
More recent methods, such as machine learning-based survival models, including survival forests and gradient boosting approaches, have been developed to address the limitations of traditional techniques.

\subsection{Proposed Pipeline} \label{bio_bank:methods:pipeline}

\begin{figure}
    \centering
    \includegraphics[width=0.67\linewidth]{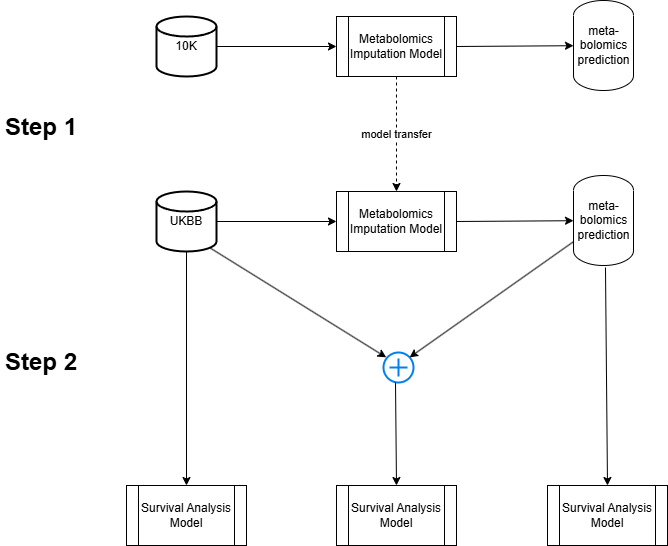}
    \caption{Proposed Pipeline}
    \label{bio_bank:fig:proposed_pipeline}
\end{figure}

In this study, we propose a two-step approach to leverage machine learning models trained on metabolomics data for disease prediction.

First, we train LightGBM models utilizing our lab database, referred to as 10K. 
These models learn to predict metabolomics features from available covariates(shared features, and nmr features separately). 
We transfer them to the UKBB dataset and apply them to impute metabolomics features for UKBB participants. 
This step allows us to extend the applicability of our models to a larger and more diverse population.

In the second step, we evaluate the utility of these imputed metabolomics features in a downstream survival analysis task. 
Specifically, we incorporate the predicted features into a Cox Proportional Hazards (CoxPH) model and compare its performance with and without them. 
This comparison enables us to assess whether metabolomics feature imputation enhances predictive performance in survival analysis.
Both ideas as described in \ref{bio_bank:fig:proposed_pipeline}.

Beyond these primary experiments, we extend our analysis in two additional directions. 
First, we identify outcomes that show improved predictive performance with the imputed metabolomics features and utilized multiple existing Genome-Wide Association Studies (GWAS-es) on the most influential features. 
This allows us to uncover related genetic traits linked to the improved outcomes. 
Second, we integrate the survival models trained on UKBB back into the 10K dataset as additional predictive inputs, allowing us to create assess the probabilities of participants to develop diseases even before growing a huge cohort containing thousands of patients.

These steps provide a comprehensive evaluation of our models, linking metabolomics predictions to survival outcomes and genetic traits while iterating improvements across datasets.

\section{Results} \label{bio_bank:results}

Our results indicate that the imputed metabolomics features were able to contribute to a deeper understanding of disease mechanisms and come up with some well-known disease factors that weren't present in their training data.
We identified genetic links between vascular dementia and smoking through GWAS and uncovered metabolic substances associated with obesity by reintegrating survival models into 10K. 
These findings highlight the potential of combining machine learning, survival analysis, and genetic studies to derive novel insights from biomedical datasets.

\subsection{Training the Models}

\begin{figure}[htbp]
\centering
    \includegraphics[scale=0.9]{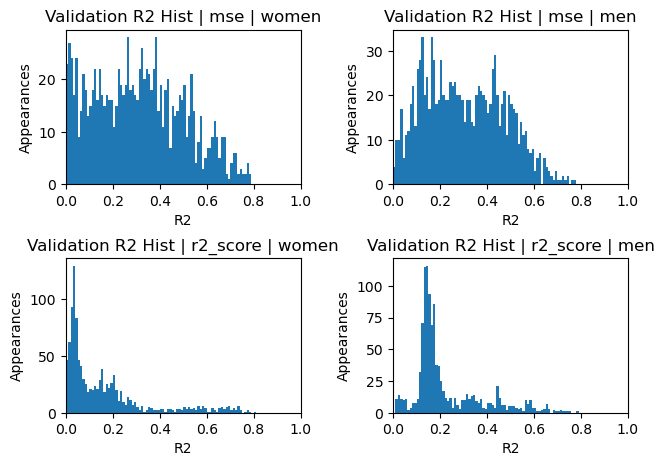}
    \caption{
        Training results for models on 10K(val set)
        \newline
        $R^2$ values for each metabolite stratified by gender, obtained using the proposed pipeline with both MSE and $R^2$-score loss functions.
    }
    \label{bio_bank:fig:train_results}
\end{figure}

We first present the results of the proposed metabolomics pipeline, as detailed in Section \ref{bio_bank:methods:pipeline}.
Figure \ref{bio_bank:fig:train_results} shows the $R^2$ values received within the training process for both mse and $R^2$ score loss functions. Both loss functions produced a substantial number of statistically significant predicted features.

\begin{figure}[htbp]
\centering
    \includegraphics[scale=0.9]{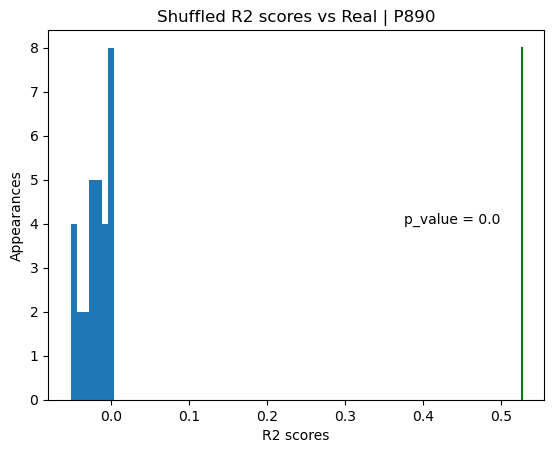}
    \caption{
        Testing significance for metabolomics columns predictions on 10K(val set)
        \newline
        Each blue columns represents an $R^2$ value of the pipeline with shuffled labels. The green columns on the right in the $R^2$ value given by the pipeline with the true labels.
    }
    \label{bio_bank:fig:metab_col_sig_test}
\end{figure}

To assess the significance of our predictions, we compared the $R^2$ scores obtained
from our proposed method with those from multiple iterations of training on shuffled labels.
Statistical significance was determined using a one-sided normal test, with the p-value computed as $\frac{Z}{2}$. This is illustrated in Figure \ref{bio_bank:fig:metab_col_sig_test}, where one example column is highlighted.
The observed p-value $(< 0.0001)$ confirms that the results obtained by the pipeline are statistically significant under the one-sided normal test $(p < 0.05)$.
All features that were found significant for both men and women were collected and kept in use for the rest of the procedure.

\subsection{Method Verifications}

After obtaining statistically significant metabolite predictions on the validation set (10K biobank), we aimed to assess the generalizability of our approach when applied to an external biobank (UKBB).
However, the UKBB dataset lacks ground-truth labels for direct validation of our predictions.
Hence we came up with two different ways of doing so.

\begin{figure}[htbp]
\centering
    \includegraphics[scale=0.48]{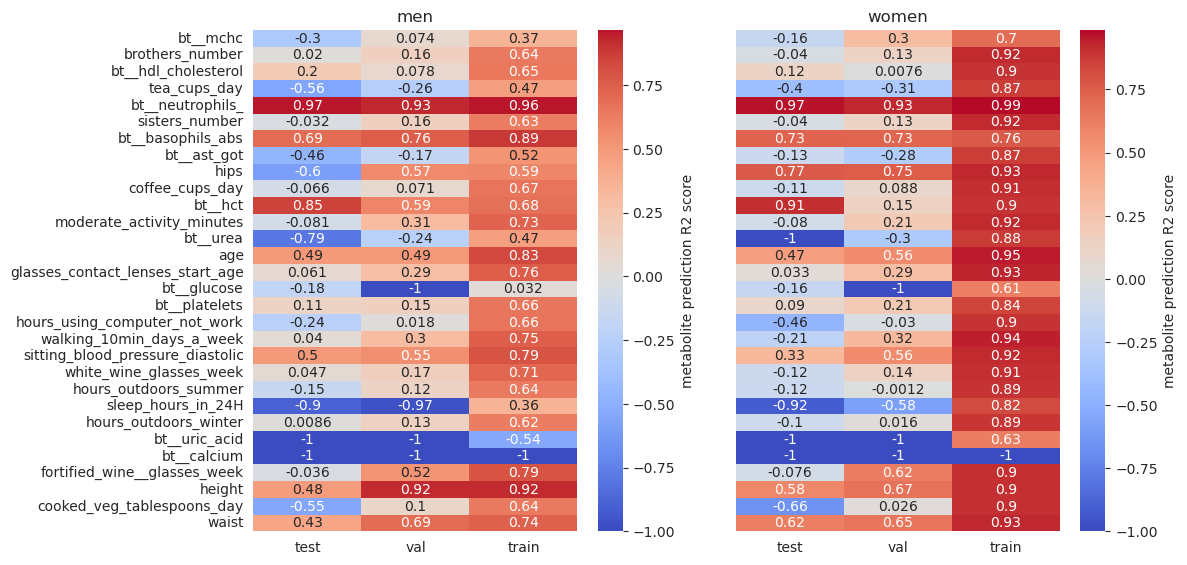}
    \caption{
        Prediction results for omitted columns in the test dataset (UKBB).
        \newline
        Both figures show the $R^2$ values given by running the main pipeline(shared features to metabolomics) when omitting all of the columns on the y-axis, and using them as labels. The columns are shows the results on the different data splits where train and test are taken from 10K, and test is taken solely from UKBB. The figures are separated by sex.
    }
    \label{bio_bank:fig:val_omit}
\end{figure}

The first way is leaving out some of the input columns, using them as labels and re-running the whole pipeline. Exploring these values gives a clue on the ability of the pipeline to transfer knowledge between the datasets. This idea is described in Figure \ref{bio_bank:fig:val_omit}. This simple test shows that the method was able to predict to good extent some of body measures and blood test features(age, waist, height, blood pressure, and so on) on the external dataset for both genders.

\begin{figure}[htbp]
    \begin{subfigure}{1.0\textwidth}
        \begin{subfigure}{0.5\textwidth}
            \includegraphics[scale=0.46]{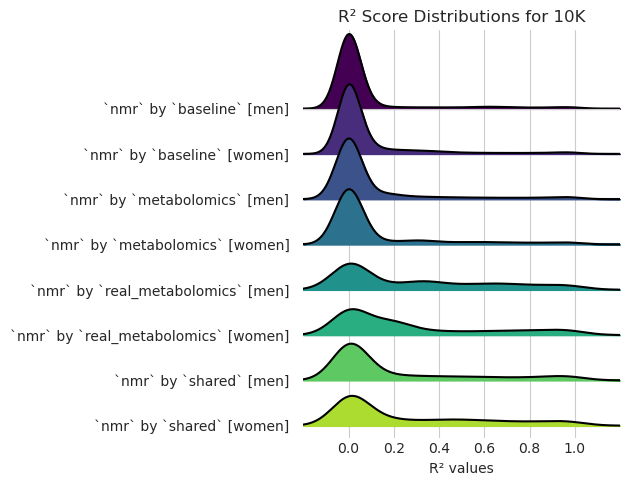}
            \caption{}
            \label{bio_bank:fig:external_modalities_shared_10k}
        \end{subfigure}
        \begin{subfigure}{0.5\textwidth}
            \includegraphics[scale=0.46]{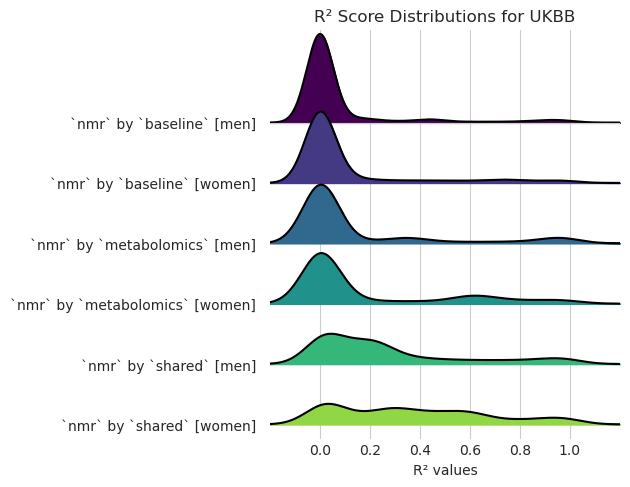}
            \caption{}
            \label{bio_bank:fig:external_modalities_shared_ukbb}
        \end{subfigure}
        \caption*{Shared features to Metabolomics}
    \end{subfigure}

    \begin{subfigure}{1.0\textwidth}
        \begin{subfigure}{0.5\textwidth}
            \includegraphics[scale=0.46]{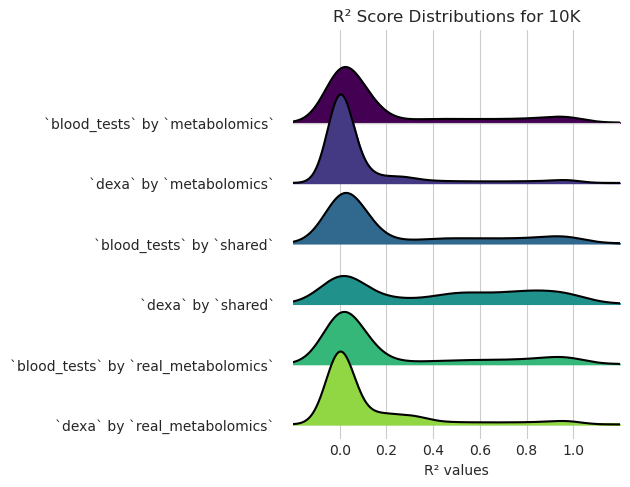}
            \caption{}
            \label{bio_bank:fig:external_modalities_nmr_10k}
        \end{subfigure}
        \begin{subfigure}{0.5\textwidth}
            \includegraphics[scale=0.46]{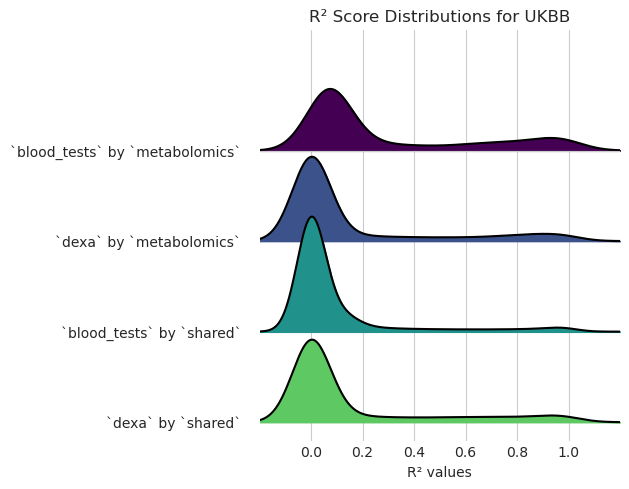}
            \caption{}
            \label{bio_bank:fig:external_modalities_nmr_ukbb}
        \end{subfigure}
        \caption*{Nmr to Metabolomics}
    \end{subfigure}
\caption{
    Predicting unseen modalities from the imputed metabolomics
    \newline
    Sub-figures \ref{bio_bank:fig:external_modalities_shared_10k} and \ref{bio_bank:fig:external_modalities_shared_ukbb} shows the external modalities prediction results from the pipeline trained from the shared features, where sub-figures \ref{bio_bank:fig:external_modalities_nmr_10k} and \ref{bio_bank:fig:external_modalities_nmr_ukbb} shows the external modalities prediction results from the pipeline trained from the nmr features.
}
\label{bio_bank:fig:external_modalities}
\end{figure}

Although this method indicated a potential signal, its effectiveness was constrained by
the concentration of relevant information in a small subset of features. Consequently,
it exhibited sensitivity to feature removal, limiting its predictive power to groups of features.
Therefore we present another way of verifying our proposed pipeline - predicting external unseen modalities.
This is illustrated in Figure \ref{bio_bank:fig:external_modalities}.
Both pipelines demonstrated strong predictive performance across the 10K and UKBB datasets.
The shared features to metabolomics pipeline succeeded at the task of predicting nmr from both metabolomics and the shared features separately(shown at \ref{bio_bank:fig:external_modalities_shared_10k} and \ref{bio_bank:fig:external_modalities_shared_ukbb}).
The nmr to metabolomics pipeline succeeded at the tasks of predicting blood tests results and dexa results from both metabolomics and the shared features separately(shown at \ref{bio_bank:fig:external_modalities_nmr_10k} and \ref{bio_bank:fig:external_modalities_nmr_ukbb}).

\subsection{Outcome Predictions}

Having a validated method that predicts metabolomics, allows for the imputation of this modality.
We next evaluate the imputed metabolomics features in an external biobank containing labeled outcomes.
This opens the door for outcome prediction using survival analysis methods.

\begin{figure}[H]
    \begin{subfigure}{0.5\textwidth}
        \includegraphics[scale=0.47]{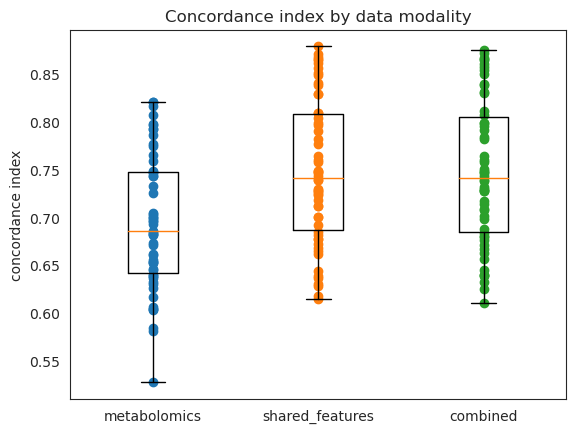}
        \caption{}
        \label{bio_bank:fig:outcomes_shared}
    \end{subfigure}
    \begin{subfigure}{0.5\textwidth}
        \includegraphics[scale=0.47]{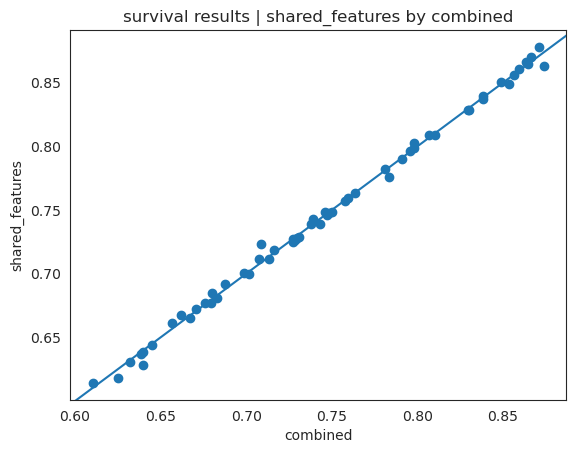}
        \caption{}
        \label{bio_bank:fig:outcomes_reg_shared}
    \end{subfigure}
    
    \begin{subfigure}{0.5\textwidth}
        \includegraphics[scale=0.47]{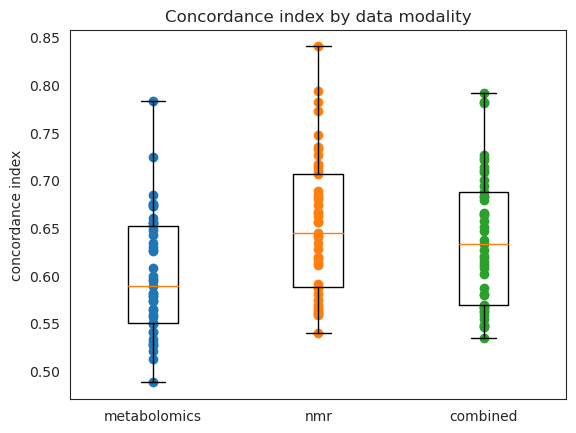}
        \caption{}
        \label{bio_bank:fig:outcomes_nmr}
    \end{subfigure}
    \begin{subfigure}{0.5\textwidth}
        \includegraphics[scale=0.47]{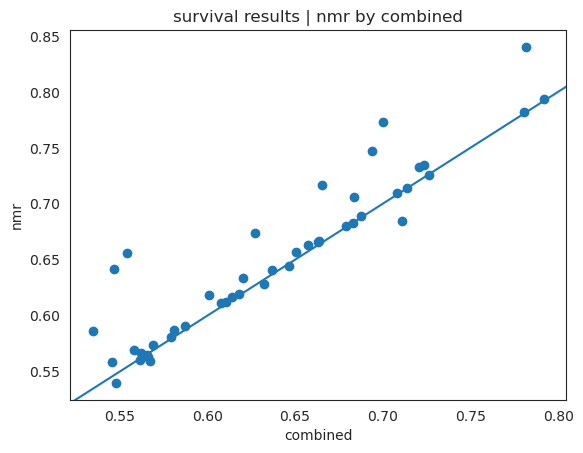}
        \caption{}
        \label{bio_bank:fig:outcomes_reg_nmr}
    \end{subfigure}  
    
    \caption{
        Comparison of outcome predictions across different modalities.
        \newline
        Plots (a) and (c) contains outcome predictions given from models based on shared features (a) and nmr (c) modalities. Plots (b) and (d) shows the side by side analysis per each outcome for the different modalities.
    }
    \label{bio_bank:fig:outcome_preds}

\end{figure}

We used CoxPH models on survival analysis tasks per each outcome, for each modality, namely shared input features, predicted metabolomics features, and all features combined. We then repeated this for both pipelines - trained from the shared features(body measures, blood tests, and lifestyle questionnaires), and the nmr metabolomics respectively. This is shown at Figure \ref{bio_bank:fig:outcome_preds}.

We tested the combined outcome results against the base input features for the pipelines. 
The combined shared-features-based pipeline demonstrated competitive performance but did not yield statistically significant improvements in outcome prediction, as results remained highly correlated with baseline features.
The combined nmr metabolomics based pipeline fell short of the performance of the original nmr features by themselves, yet was able to improve some of the outcomes predictions. Annotated examples could be illustrated at \ref{bio_bank:fig:explain_candidates} displaying the improved predictions for vascular dementia, hyperthyroidism and inflammatory liver diseases.

\begin{figure}[htbp]
    \centering
    \begin{subfigure}{1.0\textwidth}
        \centering
        \includegraphics[scale=0.5]{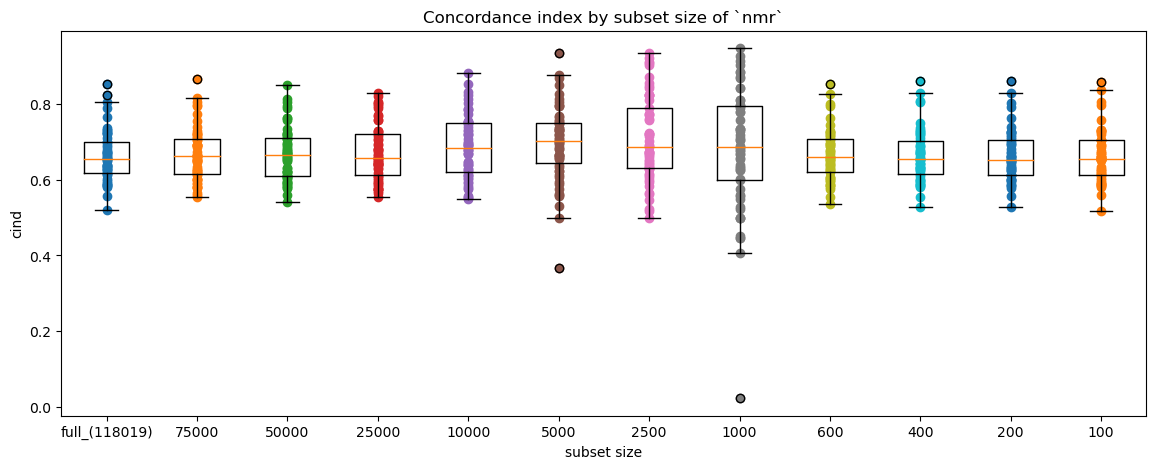}
        \caption{}
        \label{bio_bank:fig:nmr_subsets}
    \end{subfigure}
    
    \begin{subfigure}{1.0\textwidth}
        \centering
        \includegraphics[scale=0.5]{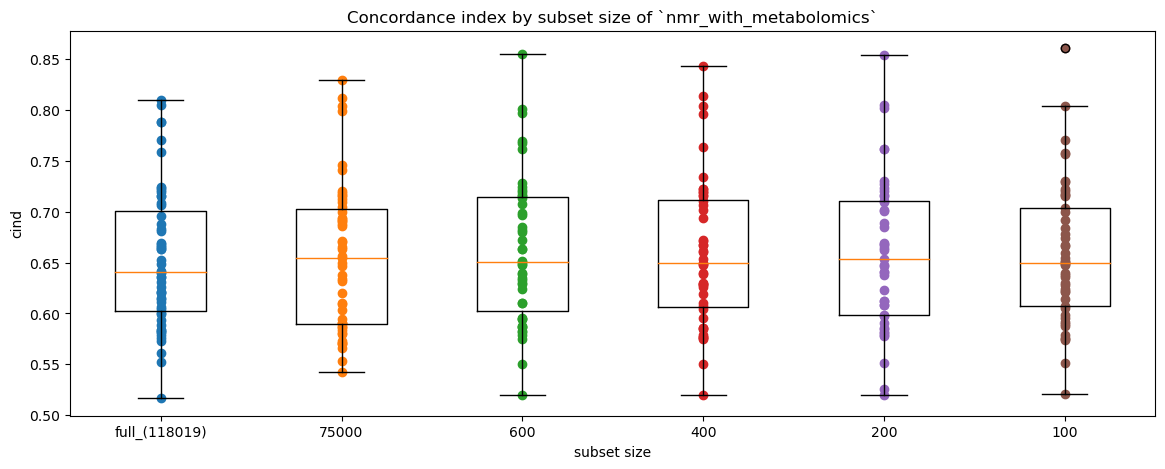}
        \caption{}
        \label{bio_bank:fig:combined_subsets}
    \end{subfigure}
    
    \begin{subfigure}{1.0\textwidth}
        \centering
        \includegraphics[scale=0.6]{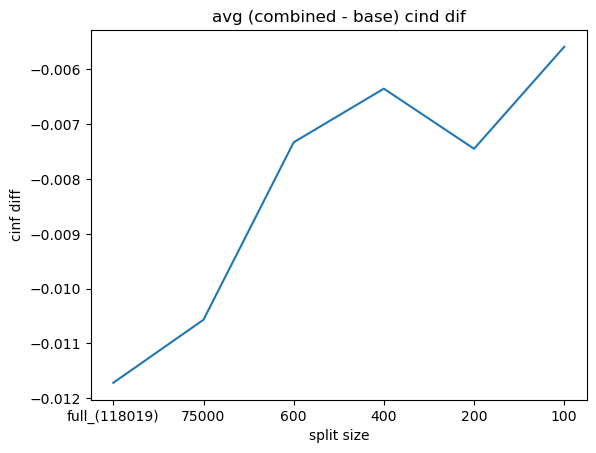}
        \caption{}
        \label{bio_bank:fig:subsets_cind_diff}
    \end{subfigure}

    \caption{Running the survival analysis from a subset of the data of a chosen size. The rest of the data is used for test. Sub-figures (a) and (b) shows the performance of the outcome analysis given by decreasing the input size. Sub-Figure (c) shows the difference in the performance between a model train on the nmr data, and a model trained on the combined data.}
    \label{bio_bank:fig:survival_subsets}
\end{figure}

Further exploring the method's limits we kept on to gradually decrease the size of the training set for the survival analysis task, and see whether the method could out perform in any such case. 
This can shown in details in \ref{bio_bank:fig:survival_subsets}. 
The main conclusion seen from \ref{bio_bank:fig:nmr_subsets} and \ref{bio_bank:fig:combined_subsets} is that the performances of both method weren't affected by the change in the training set size. This conclusion is also backed in Figure \ref{bio_bank:fig:subsets_cind_diff} showing that in average the method wasn't able to improve on the outcome prediction task.

\subsection{Explaining Outcomes}

Upon having outcomes predicting models, we can use them in various ways to improve our understanding on the outcome.

For all survival outcome prediction tasks where the corresponding combined NMR + Metabolite model outperformed the base NMR model for the same outcome, we identified the metabolites with highest regression coefficients in the combined model. We identified significant genome-wide associations for these metabolites using GWAS, and mapped the significant hits to genes. We queried the GWAS catalog \cite{gwas_cat} for these genes in a Phenome-Wide Association Study (PheWAS), identifying traits with similar genetic architecture to those metabolites. This yielded insight into the shared genetic architecture between not just the survival outcomes we tried to predict, but also their metabolic basis, and other traits measured in our study and beyond.

\begin{figure}[htbp]
    \begin{subfigure}{0.5\textwidth}
        \includegraphics[scale=0.48]{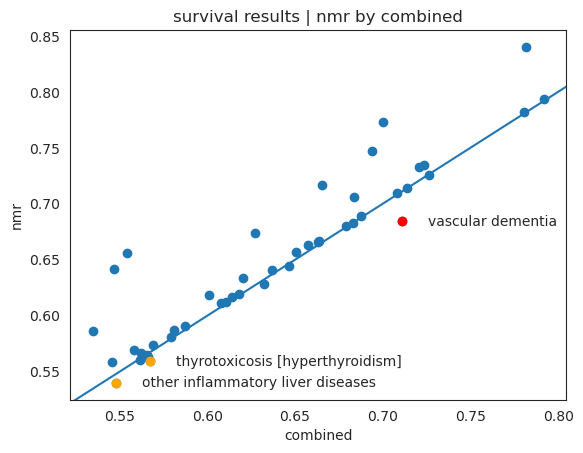}
        \caption{}
        \label{bio_bank:fig:explain_candidates}
    \end{subfigure}
    \begin{subfigure}{0.5\textwidth}
        \includegraphics[scale=0.36]{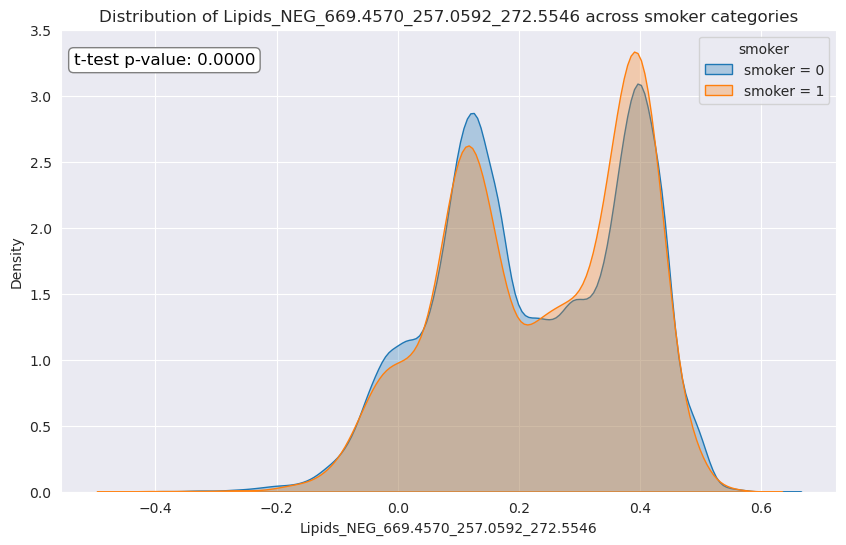}
        \caption{}
        \label{bio_bank:fig:explain_genetic_cor}
    \end{subfigure}

    \caption{
        Explainability from genetic correlation. 
        \newline
        Plots (a) displays the outcomes the proposed method was able to improve. These candidates served as inputs to multiple GWAS-s to discover genetic correlation between the outcomes and new data. One example of it is the link between the vascular dementia and smoking, in (b).
    }
    \label{bio_bank:fig:explain}
\end{figure}

To further validate the discovered outcome-related trait, we analyzed the original metabolite used in the GWAS-derived connection, testing its distribution difference by the discovered trait with t-test.
Figure \ref{bio_bank:fig:explain_genetic_cor} shows such a case between a metabolite and smoking status achieving p-value $<0.0005$.
Further verification is drawn from \cite{dementia_smoking, dementia_smoking_2}.

\subsection{Exploring Outcomes Time Effects}

\begin{figure}[htbp]
    \begin{subfigure}{0.5\textwidth}
        \includegraphics[scale=0.46]{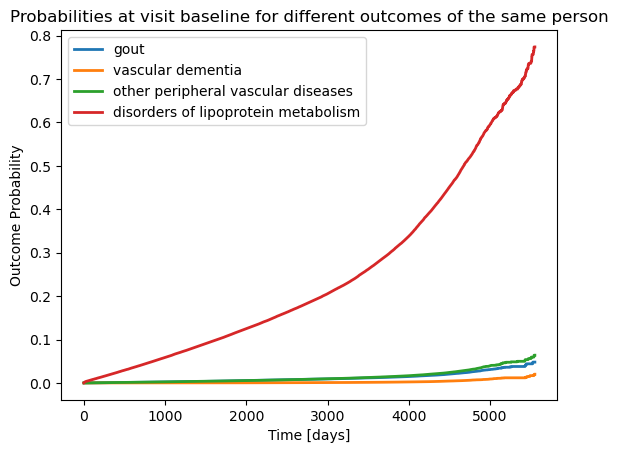}
        \caption{}
        \label{bio_bank:fig:labels_outcomes}
    \end{subfigure}
    \begin{subfigure}{0.5\textwidth}
        \includegraphics[scale=0.46]{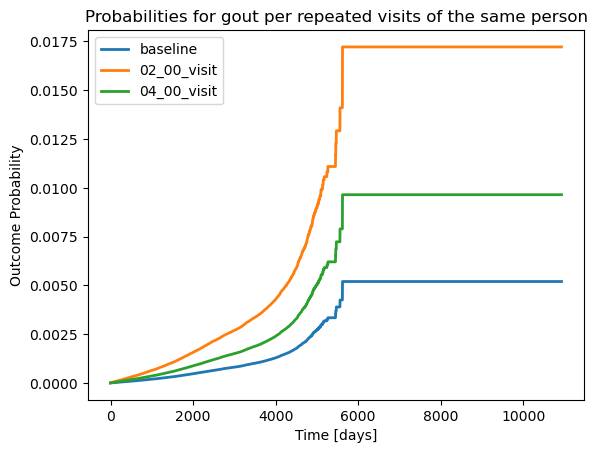}
        \caption{}
        \label{bio_bank:fig:labels_visits}
    \end{subfigure}

    \caption{
        Survival analysis with the trained models on external data.
        \newline
        Sub-Figure (a) shows a few selected outcomes plots for the same person over time, where Sub-Figure (b) shows a the same outcome function - gout, by multiple visits of the same person.
    }
    \label{bio_bank:fig:labels}
\end{figure}

We further investigate the predicted outcomes by applying an alternative analytical approach.
Returning to the 10K dataset, we leverage multiple repeated visits at fixed time intervals for many patients, though without labeled outcomes.
We utilized our survival analysis models to predict the time survival curves of the outcomes trained against for each 10K patient, this creating labels.
Survival functions as in \ref{bio_bank:fig:labels}, are the function of an event to occur throughout time. We flipped the the y axis for convenience, used it to plot the affect of the time on the outcomes.

We start with calculating partial correlations between our metabolomics features and the outcome-visit pair for a few selected outcomes we believed could be influenced by metabolomics(metabolic disorders, hypertension, gout, obesity, etc). Subsequently, only statistically significant correlations, as determined by a t-test, are selected.
We then focus the most frequent significant metabolites, which are unannotated metabolomics substances, and joined them based on mz(mass to charge ratio) and ccs(collision cross section) with a table of substances given from \cite{metabolon} to get annotated substances.

\begin{figure}[htbp]
    \centering
    \includegraphics[scale=0.5]{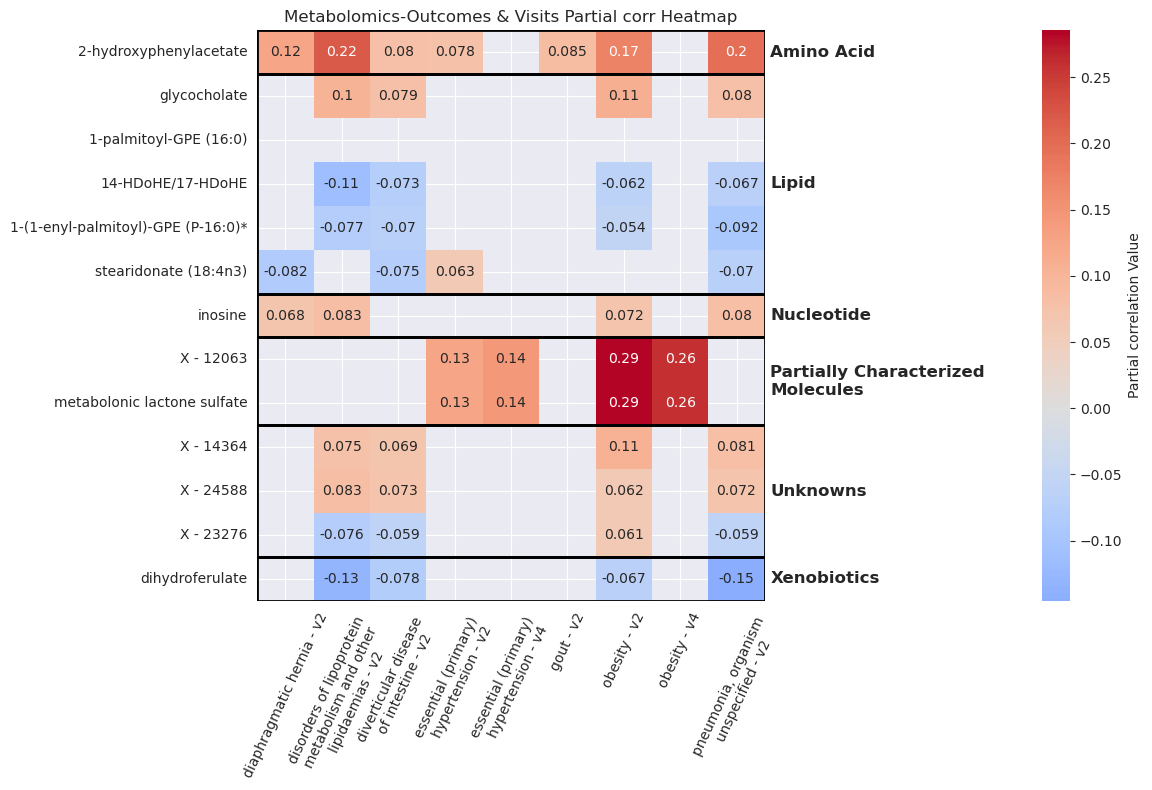}
    \caption{
        Partial Correlation of metabolites and outcome-visit pairs
    }
    \label{bio_bank:fig:visits_heatmap}
\end{figure}

\noindent
Figure \ref{bio_bank:fig:visits_heatmap} illustrates strong correlations between specific outcomes.
Some interesting one are `X-12063` and `metabolic lactone sulfate`, both very probably annotations of the same raw metabolic substance, correlated with obesity. The link between `X-12063` and obesity has already been found in \cite{hers_phd}. The link between `metabolic lactone sulfate` and obesity could be verified from \cite{solphat_nature}.

\section{Discussion}
\verify[Short interpretation of the results]

The findings of this study highlight the potential of leveraging machine learning to integrate external bio-bank data for disease prediction and biomedical discovery. 
By training LightGBM models on a smaller but metabolomics-rich dataset (10K) and applying them to the UK Biobank (UKBB), we successfully imputed missing metabolomics features and evaluated their impact on downstream survival analysis. 
Our results indicate that while the imputation process uncovered biologically relevant associations, it did not consistently enhance disease outcome predictions. 
This suggests that while imputed data can reveal meaningful biological patterns, its utility for direct risk estimation may be limited.

A key strength of our approach was its ability to extract biologically relevant insights by combining machine learning-based imputation with survival analysis and genome-wide association studies (GWAS). 
The integration of GWAS validated the biomedical significance of the imputed features, as demonstrated by the identification of a genetic link between vascular dementia and smoking—an association that was not embedded in the original training data. 
Furthermore, leveraging survival models trained on UKBB back into 10K enabled the identification of metabolic substances associated with obesity, reinforcing the potential of external biobank data for uncovering novel disease mechanisms.

\section{Limitations}
\verify[
What isn't covered in the research
(stuff we didn't aim to solve)
The method wasn't particularly successful in improving outcome prediction/survival analysis.
]

While our imputation approach successfully identified meaningful biological associations, it did not consistently improve disease outcome predictions in survival analysis. 
This suggests that although the imputed metabolomics features capture latent biological signals, many disease risks are already well-explained by existing clinical and demographic variables, limiting the additional predictive power of the imputed data.

A crucial factor enabling the transferability of our approach is the population similarity between 10K and UKBB. 
Both biobanks share overlapping data modalities and demographic characteristics as shown in \cite{levine}, allowing the imputation models to generalize reasonably well. 
However, this may not hold when extending the method to biobanks with different population structures, health conditions, or data collection methodologies. 
Ensuring that imputation models can adapt to more diverse datasets remains an open challenge that requires further investigation.

Another challenge is the difficulty in directly validating imputed features in the UKBB dataset. 
Since UKBB does not provide ground-truth labels for the imputed metabolomics features, it is challenging to assess their accuracy quantitatively. 
Our validation relied on indirect methods, such as assessing whether known biological associations emerged in downstream analyses, but a more direct evaluation of imputation accuracy remains an area for future research.

\section{Future Work}
\verify[
What isn't covered in the research
(stuff we didn't aim to solve)
The method wasn't particularly successful in improving outcome prediction/survival analysis.
Concrete thing - Use var autoencoders fit to predict metabolomics (as in one of the citations)
]

To improve the utility of imputed metabolomics features, future research should explore more advanced imputation techniques. 
Deep generative models, such as variational autoencoders (VAEs), may enhance feature reconstruction by better capturing nonlinear dependencies in biological data. 
Additionally, incorporating uncertainty estimation in imputation models could provide more reliable predictions, particularly when handling missing data that is not random.

Integrating multiple data modalities could further enhance both imputation and disease prediction. 
Beyond metabolomics, incorporating proteomics, transcriptomics, or imaging data may offer a more comprehensive understanding of disease risk. 
Multi-modal learning approaches, such as transformer-based architectures or multi-modal VAEs, could improve imputation accuracy while uncovering deeper biological insights.

Expanding the validation of imputation models across additional biobanks is also crucial. 
Testing the approach on datasets with varying demographic compositions and disease prevalence rates will help assess its robustness and generalizability. 
Additionally, integrating causal inference techniques with imputation could help distinguish between merely correlated features and those with direct mechanistic relevance to disease progression.

By refining imputation methodologies, incorporating multi-modal data, and validating across diverse cohorts, future research can further leverage external biobanks to enhance disease modeling, ultimately contributing to improved biomedical insights and precision medicine.

\section{Conclusion}
\verify[
The results achieved in the paper, against the shortly description in the abstract
]

This study demonstrates the feasibility of leveraging external biobanks to enhance disease modeling through metabolomics imputation. 
While direct improvements in outcome prediction were limited, our approach successfully identified biologically relevant associations and reinforced known epidemiological links. 
The integration of GWAS validated the biomedical significance of imputed features, as demonstrated by uncovering the genetic link between vascular dementia and smoking. 
Additionally, incorporating survival models from UKBB into 10K facilitated the identification of metabolic pathways associated with obesity, showcasing the broader potential of external biobank data for biomedical discovery.

These findings highlight both the opportunities and challenges of using machine learning for biomedical data integration. 
The results emphasize the need for more sophisticated imputation techniques and multi-modal data integration to enhance predictive performance. 
Future advancements in generative modeling, validation across diverse populations, and causal inference will be critical to fully realizing the potential of machine learning in metabolomics-driven disease prediction.

\end{document}